%12-09-94 including comments from ref of 12-07-94
\input harvmac
\def\yesanswfig{y }
\def\answfig{y }
\ifx\answfig\yesanswfig\message{Figures will be included 
automatically using epsf.tex}
\def\INSERTFIG#1#2{\vbox{\hfil\epsfbox{#1}\hfill}#2}\def\INSERTCAP#1{}
\else\message{Figures will not be included automatically.
Figure captions will be appended at end}\def\INSERTFIG#1{}
\def\INSERTCAP#1{#1}
\fi
%
%% Figures:  There are 5 postscript figures packed with "uufiles,"
%%
%% If you don't want them embedded,
%% uncomment the next line and delete the one after that.
%
% % uncomment to stop figure insertion
\input epsf

\def\CAPTION#1#2{{\narrower\noindent
\multiply\baselineskip by 3
\divide\baselineskip by 4
{\ninerm #1}{\ninesl #2 \medskip}}}
%
%from btok
\def\etal{{\it et al.}}
\def\b{{\rm b}}
\def\s{{\rm s}}
\def\ol{\overline}

\def\O{{\cal O}}

\def\d{{\rm d}}
\def\im{{\rm i}}
\def\q{{\rm q}}
\def\vslash{v\hskip-0.5em /}
\def\Aslash{A\hskip-0.5em /}

\def\OMIT#1{}
\def\frac#1#2{{#1\over#2}}
\def\eps{{\epsilon}}
\def\pintegral{\int{\d^{4-\eps}\,p\over(2\pi)^{4-\eps}}}

%
%my own
\def\cM{ {\it \Pi}}
\def\mM{ {m_K^2 \over 16 \pi^2 f_K^2 M}}
\def\mg{ {m_K^2 g^2 \over 16 \pi^2 f_K^2}}
\def\dbar{\overline \Delta}
\def\ordone{${\cal O}(1)$}
\def\ord{${\cal O}({1\over M}$)}
\def\orms{${\cal O}({1\over M^2}$)}
\def\orlm{$\O({\lbar \over M})$}
\def\ro{\rho_1}
\def\rt{\rho_2}
\def\Lzst{ {m_K^2( \ln\frac{m_K^2}{\mu^2}+ \frac23+\frac4{13} 
\ln \frac43) \over 16 \pi^2 f_K^2}\,}
\def\Lz{ {m_K^2( \ln\frac{m_K^2}{\mu^2}+ \frac23+\frac2{11} 
\ln \frac43) \over 16 \pi^2 f_K^2}\,}
\def\Lzrone{ {m_K^2( \ln\frac{m_K^2}{\mu^2}+ \frac23+\frac2{5} 
\ln \frac43) \over 16 \pi^2 f_K^2}\,}
\def\L{ {m_K^2 \ln\frac{m_K^2}{\mu^2} \over 16 \pi^2 f_K^2}\,}
\def\LMz{ {m_K^2(\ln\frac{m_K^2}{\mu^2}+ \frac23+\frac2{11} \ln 
\frac43) \over 16 \pi^2 f_K^2 M}\,}
\def\LMzst{ {m_K^2(\ln\frac{m_K^2}{\mu^2}+ \frac23+\frac4{13} 
\ln \frac43) \over 16 \pi^2 f_K^2 M}\,}
\def\LM{ {m_K^2 \ln\frac{m_K^2}{\mu^2} \over 16 \pi^2 f_K^2 M}\,}

\def\Ddp#1{\Delta^{(#1)} + \delta}
\def\Ddm#1{\Delta^{(#1)} - \delta}
\def\fcplus{ F({m_K \over \Delta^{(D)} + \delta})  }
\def\fcminus{ F({m_K \over \Delta^{(D)} - \delta})  }
\def\fbminus{ F({m_K \over \Delta^{(B)} - \delta})  }
\def\fbplus{ F({m_K \over \Delta^{(B)} + \delta})  }
\def\Lfph{ {(\delta+ \Delta)^2 \hat g^2 \over \pi^2 f_K^2}
     \[\ln\frac{m_K^2}{\mu^2} + 2 F({m_K \over \Delta + \delta}) \] \,}
\def\Lfp{ {(\delta+ \Delta)^2 g^2 \over \pi^2 f_K^2}
     \[\ln\frac{m_K^2}{\mu^2} - \frac13 + 2 F({m_K \over \Delta +
	\delta}) \] \,}
\def\Lfm{ {(\delta-\Delta)^2 g^2 \over \pi^2 f_K^2}
     \[\ln\frac{m_K^2}{\mu^2} - \frac13 + 2 F({m_K \over \Delta -
	\delta}) \] \,}
\def\lbar{\bar \Lambda}
\def\gt{\tilde g}

\def\etal{{\it et al.} }
\def\dfour{\frac{\Delta}4}
\def\dmthree{\frac{-3 \Delta}4}
\def\vk{v \cdot k}
\def\vt{\tilde v}
\def\xt{\tilde x}
%
% From rad
\def\np#1#2#3{{Nucl.~Phys.}~B{#1} (#2) #3}
\def\pl#1#2#3{{Phys.~Lett.}~{#1}B (#2) #3}
\def\prl#1#2#3{{Phys.~Rev. Lett.}~{#1} (#2) #3}
\def\physrev#1#2#3{{Phys.~Rev.}~{#1} (#2) #3}

\def\del{\partial}

\def\bar#1{\overline{#1}}

\def\Tr{\mathop{\rm Tr}}

\def\GeV{{\rm GeV}}
\def\MeV{{\rm MeV}}

%
   %curly letters

   \def\CL{{\cal L}}
\def\CM{{\cal M}}  \def\CO{{\cal O}}

\def\[{\left[}
\def\]{\right]}
\def\({\left(}
\def\){\right)}

\def\chiral{$SU(3)_L\times SU(3)_R$}
\def\sproj{{1\over 2} \(1+\vslash\)}

\def\chiral{SU(3)_L\times SU(3)_R}
\def\gmu{\gamma_\mu}

\def\[{\left[}
\def\]{\right]}
\def\gamf{\gamma_5}

\Title{\vbox{\hbox{UCSD/PTH 93-46}
\hbox{SMU-HEP/94-03}\hbox{SSCL--Preprint--532}
\hbox{hep-ph/9402340}}}
{Chiral and Heavy Quark Symmetry Violation in B Decays}

\centerline{C. Glenn Boyd and Benjam\'\i n Grinstein}
{\it\centerline{Dept. of Physics 0319, University of
California at San Diego}
\centerline{9500 Gilman Dr, La Jolla, CA 92093}}

\vskip .3in
The most general Lagrangian consistent
with chiral, heavy quark, and strong interaction symmetries
to order $\frac1M$ and to linear order in the $SU(3)$ vector
and axial currents is presented.
Two new dimensionful and five dimensionless couplings arise at
this order. The heavy to light flavor changing current is
derived to the same order, giving rise to two additional
dimensionful constants and six dimensionless ones. The 
dimensionless parameters are shown to be irrelevant at \ord.
Analytic SU(3) counterterms are also considered.  
Form factors for $D \to \pi l \nu$ and 
$\ol B \to \pi l \ol \nu$ are computed at $\CO(\frac1M,m_s^0)$
and $\CO(M^0,m_s^1)$. The eight decay constants $f_D$,
$f_{D_s}$, $f_{D^*}$, $f_{D_s^*}$,
$f_B$, $ f_{B_s}$, $f_{B^*}$ and $f_{B_s^*}$ are computed
at  $\CO(\frac1M,m_s)$ in terms of seven parameters which 
can be determined by $\{\ol B,D\} \to \{\ol K, \pi\} l \ol \nu$ decays, 
and two undetermined counterterms.  
The ratio $R_1 = \frac{f_{B_s}}{f_B}/\frac{f_{D_s}}{f_D}$
is expressed in terms of four parameters. 

% \draft
\Date{February 1994; Revised November 1994}
\vfil\eject
I. INTRODUCTION

The incorporation of chiral and heavy quark symmetries into a
Lagrangian governing the interactions of heavy mesons with
pseudo-Nambu-Goldstone bosons leads to new relations
among heavy meson leptonic and semi-leptonic form
factors. Because the charm quark mass is not large, and
the strange mass not small, one expects significant
deviations from the heavy quark and chiral limits in
many processes.  It is therefore important to compute these
corrections both to improve the accuracy of theoretical
relations and to test the validity of the perturbative
expansion.

SU(3) corrections to heavy meson decay constants
have been found to be around $20\%$ in a chiral log
approximation\ref\fchiralcalc{J. Goity, \physrev{D46}{1992}{3929}\semi
B. Grinstein, \etal, \np{380}{1992}{369}.}.
\nref\bernard{C. Bernard, J. Labrenz, and A. Soni,
\physrev{D49}{2536}{1994}}%
\nref\fhqcalc{B. Grinstein and P. Mende, \prl{69}{1992}{1018}.}%
\nref\neubert{M. Neubert, \physrev{D46}{1992}{1076}.}%
There is suggestive evidence that there are, in addition, large heavy quark
symmetry violations\refs{\bernard{--}\neubert}, although
the question remains open \ref\virial{M. Neubert, \pl{B322}{419}{1994}.}.
However, the leading corrections to decay constants which
violate both chiral and heavy quark symmetries have
not, to our knowledge, been examined. That is the primary purpose of this
paper.

Quantities which are sensitive only to violations of both
symmetries, such as the ratio
\eqn\ratiodef{
        R_1 = \frac{f_{B_s}}{f_B}/\frac{f_{D_s}}{f_D}, }
can be predicted with greater accuracy\ref\ratio{B. Grinstein, 
\prl{71}{3067}{1993}.}, and can provide
information on $\frac{\lbar}{M_B}$ and
$\frac{m_s}{\Lambda_\chi}$ coupling constants.  These couplings also
enter into
\nref\bunch{J.Goity, \physrev {D46}{1992}{3929}\semi
J.Rosner and M.Wise, \physrev {D47}{1992}{343}\semi
L.Randall and E. Sather, \pl {303}{1993}{345}}
\nref\jenkins{E. Jenkins, \np{B412}{181}{1994}}
such processes as $B \to D$ and
$B \to K$ semileptonic decays, $B-\ol B$ mixing, and
hyperfine mass splittings\refs{\bunch,\jenkins}.

To see how $R_1$ deviates from unity requires the heavy quark chiral
Lagrangian to \ord. We derive both the Lagrangian and
the heavy-light current to this order in the next section. We also
present analytic counterterms linear in the light quark mass matrix.
Implications for $B \to \pi l \nu$ decays are then considered, with
emphasis on separately heavy quark or chiral symmetry violating
quantities. The remainder of the paper is
devoted to a computation of the leading heavy quark and chiral
symmetry violating corrections to heavy meson decay constants,
including both analytic and nonanalytic contributions.  

The low momentum strong interactions of $B$ and $B^*$ mesons
are governed by the chiral Lagrangian
\ref\wbdy{M.~Wise, \physrev{D45}{1992}{2188}\semi
G.~Burdman and J.~F.~Donoghue, \pl{280}{1992}{287}\semi
T.~M.~Yan \etal, \physrev{D46}{1992}{1148}}
\eqn\original{\eqalign{
       {\cal L}&=
    -\Tr\left[\overline H_a(v)\im v\cdot D_{ba} H_b(v)\right]\cr
    &+ g\,\Tr\left[\overline H_a(v)H_b(v)\,\Aslash_{ba}\gamma_5\right]\,. }}
Operators suppressed by powers of the heavy meson mass\foot{
Unless the deviation
of the expansion mass from $M_B$ is suppressed by $\frac1M$,
the Lagrangian \original\ must include an explicit mass term.
We choose the spin-weighted average of
the $B$ and $B^*$ masses as our expansion parameter, allowing
us to use the equation of motion $v \cdot k \sim \frac1{M^2}$,
where $k$ is the residual momentum of the meson.}
$1/M_B$, factors of
a light quark mass $m_q$, or additional derivatives have been
omitted.
The field $\xi$ contains  the octet of pseudo-Nambu-Goldstone bosons
\eqn\pseudo{\xi=\exp\(i\cM/f\),}
where
\eqn\cm{\cM=\pmatrix{{\textstyle{1\over\sqrt{2}}}\pi^0+{\textstyle
    {1\over\sqrt{6}}}\eta&\pi^+&K^+\cr\pi^-
    &{\textstyle{-{1\over\sqrt{2}}}}\pi^0+
    {\textstyle{1\over\sqrt{6}}}\eta&K^0\cr K^-&\bar K^0&-{\textstyle\sqrt
    {2\over 3}}\eta}.}
The bosons couple to
heavy fields through the covariant derivative and the axial vector field,
\eqn\covax{
    \eqalign{&D_{ab}^\mu=\delta_{ab}\del^\mu+V_{ab}^\mu
    =\delta_{ab}\del^\mu+\half\left(\xi^\dagger
    \del^\mu\xi+\xi\del^\mu\xi^\dagger\right)_{ab}\,,\cr
    &A_{ab}^\mu={\im \over 2}\left(\xi^\dagger\del^\mu\xi
    -\xi\del^\mu\xi^\dagger\right)_{ab}
    =-{1\over f}\partial_\mu{\it \Pi}_{ab}+{\cal O}({\it \Pi}^3)\,.}}
Both the vector and axial vector fields are traceless, $A^\mu_{aa} = 0 =
V^\mu_{aa}$.  Since $f_\pi$ and $f_K$ are equivalent at tree
level, we will use $f_\pi\simeq 132\,\MeV$
when pions are involved, and $f_K\simeq 161\,\MeV$ otherwise.

Under chiral $\chiral$ transformations,
\eqn\transf{\xi\to L\xi U^\dagger = U\xi R^\dagger,}
where $L$ and $R$ are elements of $SU(3)_L$ and $SU(3)_R$,
respectively, and $U$ is defined implicitly by Eq.~\transf.
The $B$ and $B^*$ heavy meson fields are incorporated into
the $4 \times 4$ matrix $H_a$:
\eqn\defineh{\eqalign{H_a&=\sproj\[\ol B^{*\mu}_a\gmu - \ol B_a\gamf\],\cr
             \bar H_a&=\gamma^0 H_a^\dagger \gamma^0\,.}}
The four-velocity of the heavy meson is $v^\mu$, and
the index $a$ runs over light quark flavor. The bar over $B$
will sometimes be omitted for notational simplicity.
% so that $(\ol B_1,\ol B_2,\ol B_3)= (\bbminus,\bbz,\bbs)$
%and $(\ol B_1^*,\ol B_2^*,\ol B_3^*)=(\bbms,\bbzs,\bbss)$.
Under $SU(2)_v$
heavy quark spin symmetry and chiral $\chiral$ symmetry, $H_a$ transforms
as
\eqn\trans{H_a\to S(HU^\dagger)_a\,,}
where $S\in SU(2)_v$. The covariant derivative has simple
transformation properties under chiral $\chiral$:
\eqn\DH{
    D\ol H \to U( D\ol H) ,}
\eqn\Dxi{
    D\xi \to U ( D\xi) R^\dagger}
and
\eqn\Dxidagger{
    D\xi^\dagger \to U ( D\xi^\dagger) L^\dagger.}

The left handed current is represented by
  \eqn\current{
   J_a^\lambda =  \frac{i\alpha}2 \Tr[\Gamma H_b(v)\xi^\dagger_{ba}] }
where $\Gamma = \gamma^\lambda (1 - \gamma_5)$.
Tree level matching to the definition of the decay constant
$f_B$ gives $\alpha = \sqrt{M_B} f_B$.

All formulas hold for the D meson as well, after
the substitution $M_B,f_B \to M_D, f_D$.
Thus the axial coupling constant $g$ is responsible for $D^*\to D\pi$
decays.

\smallskip
\smallskip
II. 1/M CORRECTIONS
\smallskip

To go beyond leading order, we make some approximations based on the
formal hierarchy $m_\pi \ll m_K \ll \lbar \sim \Lambda_\chi \ll M,$
where $\lbar = M_{B} - M_b$, $M$ is a heavy meson mass, and
$\Lambda_\chi$ is the chiral symmetry breaking scale.  Theoretically
this hierarchy holds arbitrarily well in the $m_{u,d}\ll m_s\to0$ and
$M\to\infty$ limits. However, the relation $\lbar \sim \Lambda_\chi$
should be taken loosely; factors of three or four will be important
numerically. Such factors can only be found by explicit calculation,
or by comparing to experiment.  Since $\Delta = M_{D^*} - M_D \sim
m_\pi$, we do not make assumptions about the size of the ratio
$m_\pi/\Delta$.

New parameters appear in the Lagrangian and current beyond leading order.
Since there are more experimental observables than parameters,
verifiable relations will exist. As a means to organize the calculation, 
we will use $\ol B \to \pi l \nu$ and its symmetry-related decays
to fix, in principle, as many of these parameters as possible, 
then describe the various 
$B$ and $D$ decay constants in terms of these parameters. Accordingly, we
will examine corrections to $\ol B \to \pi l \nu$ form factors which
violate either SU(3) or heavy quark symmetry, but not both. Since SU(3) loop
corrections have been computed elsewhere\ref\btok{A. Falk and B. Grinstein,
\np{B416}{771}{1994}.}, this task requires only the use of tree graphs.
Both loop graphs and analytic counterterms, however, enter into decay constant
calculations. For decay constants, we include terms which violate SU(3) and 
heavy quark symmetry simulataneously.  

For $m_s =0$,
operators appearing in the Lagrangian at \ord\ may be either dimension
four, with dimensionful coupling constants of order $\lbar$, or
dimension five, with additional derivatives.  As will become evident,
dimension five operators containing multiple factors of the axial or
vector fields $A$ or $V$ will be relevant to neither tree level $\ol B
\to \pi e \nu$ nor one-loop decay constant calculations, so they will be
ignored. The remaining higher derivative operators formally give
contributions to $\ol B \to \pi e \nu$ of order
${k\over\Lambda_\chi}{\bar\Lambda \over M}$, where $k$ is the pion
momentum, so they can be neglected to this order in the chiral
expansion. This approximation is best in the kinematic region $k\sim
m_\pi$.  Similar statements apply to higher derivative operators in the
current.  These operators could also enter the calculation of the decay
constants, but their \ord\ contributions vanish, as we shall see below
explicitly. Such higher derivative operators will be included in our
construction of the Lagrangian and current, but we will show that they
play no role in the final results.

\OMIT{We will make use of the formally consistent but
numerically suspect approximation $\ln \frac{\mu^2}{m_K^2} \gg 1$, where
$\mu \sim 1 \GeV$ is our subtraction scale, to ignore interaction terms
involving the light quark mass matrix. 
Such logarithmic terms are only generated by loop graphs.
The subtraction scale dependence of the loop graph chiral
logs is cancelled by Lagrangian counterterms 
containing $\ln \frac{\Lambda^2}{\mu^2}$ logarithms, where the constant
$\Lambda \sim \Lambda_\chi$  does not scale with the light
quark mass.
This chiral log approximation allows us to estimate the size of
SU(3) corrections.  It also implies predictions which can test
the validity of the approximation. However, it is important to
keep in mind that $\O(m_K^2)$ corrections may be sizeable.
In the light chiral SU(3) theory, such corrections have been
incorporated with some success\ref\luet{J. Gasser and
H. Leutwyler, Ann. of Phys. 158 (1984) 142; \np{250}{1985}{465}.}, so
one might hope that a similar program will eventually be
feasible in the heavy-light sector.}

An important restriction on the Lagrangian is
that it satisfy velocity reparametrization invariance (VRI).
To ensure a velocity reparametrization invariant
Lagrangian\ref\vri{M. Luke and A. Manohar, \pl{286}{1992}{348}.},
one must use velocities and derivatives on heavy fields in the combination
$v^\mu + \frac{\im D^\mu}{M}$. One should also use fields which transform by
only a phase under velocity reparametrization. Such a field is
\eqn\vrifield{
      {\tilde H} = H + \frac1{2M}\im D_\mu[\gamma^\mu,H] }
where $D_\mu$ is the covariant derivative.

The VRI consistent Lagrangian resulting from generalizing
Eq.~\original\  in this way is
\eqn\vrilagrangian{\eqalign{
      {\cal L}&=
 -\Tr\[\overline H_a(v)( \im v\cdot D_{ba} - \frac1{2M} D^2_{ba})
  H_b(v)\]\cr
    &-\frac2M \Tr\[\ol H_a(v) ( \im v\cdot D)^2_{ba} H_b(v)\]
      + g \Tr\[\overline H_a(v)H_b(v)\,\Aslash_{ba}\gamma_5\] \cr
  &+ \frac{g}M \Tr\[\overline H_c(v) (\im \delta_{bd}\overleftarrow
   {D^\mu}_{ac} - \im \delta_{ac}D^\mu_{bd} ) H_d(v)
   \gamma_\mu \gamma_5 \] v\cdot A_{ba} \cr
  &- \frac{g}M \Tr\[\overline H_c(v)(\im \delta_{bd} v\cdot
   \overleftarrow D_{ac} - \im \delta_{ac}v\cdot D_{bd} )
    H_d(v)\,\Aslash_{ba}\gamma_5\] . \cr }}
VRI fixes the coefficients of the $D^2$ and $(\im
\overleftarrow {D^\mu} - \im D^\mu )$ operators, but the other
operators (involving $v\cdot D$) are unconstrained because they obey VRI,
to the order we are working.

Another restriction is that the Lagrangian must be time reversal
invariant. Under time reversal, the psuedoscalar field transforms as
$B_v(x) \to B_{\vt}(-\xt)$ and the vector field transforms as
$B^{*\mu}_v(x) \to B^*_{\vt\mu}(-\xt)$, where $\xt,\vt$ are the parity
reflections of $x$ and $v$ ({\it e.g.} $\vt^\mu = v_\mu$). Although
the parameter $v$
is unchanged by time reversal, the field label in $B_v^{*\mu}$ alters
to reflect the transformed transversality equation $v \cdot B^* =0 \to
\vt \cdot B^*=0$. From this and the anti-unitary nature of time
reversal, it follows that
$$    H_v(x) \to T H_{\vt}(-\xt) T^{-1} $$
and
$$    \overline H_v(x) \to T \overline H_{\vt}(-\xt) T^{-1},  $$
where $T$ is a dirac matrix obeying $ T \gamma^\mu T^{-1} = \gamma_\mu^*$.
Since the pseudoscalars transform as $\cM(x) \to - \cM(-\xt)$,
the axial current transforms as $A^\mu(x) \to A_\mu(-\xt)$.
The condition of time reversal invariance then forbids operators such as
$\Tr\left[\overline H_a(v) \sigma^{\mu \nu}
   H_b(v)\gamma_\mu \gamma_5 A_{\nu ba}\right]$
and
$\im \Tr\left[\overline H_a(v) \sigma^{\mu \nu}
  v\cdot D_{bc}  H_c(v)\gamma_\mu \gamma_5 A_{\nu ba}\right]$
from appearing in the Lagrangian.

The most general form of the heavy meson Lagrangian subject to the above
constraints is
\eqn\origlagranem{\eqalign{
      {\cal L}_M&=
        -(1 + \frac{\epsilon_1}M )\Tr\left[
                   \overline H_a(v)\im v\cdot D_{ba} H_b(v)\right]\cr
             &+ ( g + \frac{g_1}M) \Tr\left[\overline
             H_a(v)H_b(v)\,\Aslash_{ba}\gamma_5\right] \cr
            &+ \frac{g_2}{M} \Tr\left[\overline H_a(v) \Aslash_{ba}
            \gamma_5 H_b(v)\right]\cr
          &+ \frac{\lambda_2}M \Tr\left[\overline H_a(v)
             \sigma^{\mu \nu}H_a(v) \sigma_{\mu \nu}\right]\cr
             &+ \frac{\epsilon_2}{M} \Tr\left[\overline H_a(v) \sigma^{\mu \nu}
              \im v\cdot D_{ba} H_b(v) \sigma_{\mu \nu}\right]\cr
       &- \frac{\delta_0}{M} \Tr\left[
                             \overline H_a(v)(\im D)^2_{ba} H_b(v)\right]\cr
&+ \frac{\delta_1}M \Tr\left[\overline H_c(v) (\im \delta_{bd}\overleftarrow
{D^\mu}_{ac} - \im \delta_{ac}D^\mu_{db} ) H_d(v) \gamma_\mu \gamma_5 \right]
   v\cdot A_{ba} \cr
   &+ \frac{\delta_2}{M} \Tr\left[\overline H_a(v) H_b(v) \gamma_\mu \gamma_5
\right]
     \im v\cdot D_{bc} A^\mu_{ca} \cr
 &+ \frac{\delta_3}{M} \Tr\left[\overline H_a(v) H_b(v)
                                \gamma_\mu \gamma_5 \right]
    D^\mu_{bc} \im v\cdot A_{ca} \cr
 &+ \frac{\delta_4}{M} \Tr\left[\overline H_a(v) \im v\cdot D_{cb} H_c(v)
     \Aslash_{ba} \gamma_5 \right] \cr
 &+ \frac{\delta_5}{M} \Tr\left[\overline H_a(v) \sigma^{\mu \nu}
              (\im v\cdot D)^2_{ba} H_b(v) \sigma_{\mu \nu}\right]\cr
  &+ \frac{\delta_6}{M} \Tr\left[\overline H_a(v)(\im v\cdot
   D)^2_{ba} H_b(v)\right] \cr
                             \,}}
where all couplings are taken to be real. The effect of $\lambda_2=
{-M \Delta \over 2}= -\frac{M}2 (M_{B^*}-M_B)$ is merely to shift 
the $B$ and $B^*$ propagators
to ${\im \over 2(v \cdot k + \frac34 \Delta)}$ and ${-\im (g^{\mu\nu}
- v^\mu v^\nu) \over 2(v \cdot k - \frac14 \Delta)}$, respectively, so
we will ignore the $\lambda_2$ term once we make this shift.

\OMIT{
 &+ \lambda_1 \Tr\left[\overline H_a(v)H_b(v)\right] (\xi m_q \xi +
         \xi^\dagger m_q \xi^\dagger)_{ba} \cr
     &+ \lambda_1^{'} \Tr\left[\overline H_a(v)H_a(v)\right] (\xi m_q \xi +
         \xi^\dagger m_q \xi^\dagger)_{bb} \cr}
\OMIT{
The effect of $\lambda_2=
{-M \Delta \over 2}$ is merely to shift the $B$ and $B^*$ propagators
to ${\im \over 2(v \cdot k + \frac34 \Delta)}$ and ${-\im (g^{\mu\nu}
- v^\mu v^\nu) \over 2(v \cdot k - \frac14 \Delta)}$, respectively, so
we will ignore the $\lambda_2$ term once we make this shift.
Similarly, the penultimate term involving the light quark mass matrix
$m_q = {\rm diag}[0,0,m_s]$ merely shifts $v\cdot k \to v\cdot k - \delta$
in strangeness carrying heavy meson propagators, where $\delta =
M_{D_s} - M_D = M_{B_s} - M_B + \O(\frac{\lbar^2}M) $ is the SU(3)
mass splitting. Its effects on decay constants have been calculated
previously
\ref\btok{A. Falk and B. Grinstein, 
\np{B416}{771}{1994}.}, but will be reproduced here for
completeness. The final
term is chirally symmetric, and therefore irrelevant to our discussion.}

We may take $\epsilon_1$ and $\epsilon_2$ to be zero by making a spin
and flavor dependent renormalization of the heavy meson fields and
then redefining the coupling constants to absorb the $\epsilon$
dependence. The current is then constructed using the new, properly
normalized, fields.

VRI implies $\delta_0 = \frac12$ and $\delta_1 = g$. However, we will
eventually show that all the higher derivative terms (with
coefficients $\delta_i$) contribute only at order $\frac1{M^2}$, so we
will omit these terms for now.  

For $m_s \ne 0$, the Lagrangian contains additional terms 
involving the light quark mass matrix $m_q = {\rm diag}[0,0,m_s]$.
To the order we are working, only operators linear in $m_q$ and inserted
in tree graphs are relevent. 
The SU(3) violating Lagrangian contains
\eqn\strangel{\eqalign{
\CL_m &= 
  2 \lambda_1 \Tr\left[\overline H_a(v)H_b(v)\right] \CM^+_{ba}
     + 2 \lambda_1^{'} \Tr\left[\overline H_a(v)H_a(v)\right]
   \CM^+_{bb}\cr 
    &+{g \kappa_1 \over \Lambda_\chi} \CM_{ca}^+
  \Tr\left[\overline H_a(v)H_b(v)\,\Aslash_{bc}\gamma_5\right]
       +{\im g \kappa_2 \over \Lambda_\chi} \CM_{ca}^-
\Tr\left[\overline H_a(v)H_b(v)\,\Aslash_{bc}\right]\cr
   &+ {g \kappa_3 \over \Lambda_\chi} \CM_{cc}^+  
   \Tr\left[\overline H_a(v)H_b(v)\,\Aslash_{ba}\gamma_5\right]
   + {\im g \kappa_4 \over \Lambda_\chi} \CM_{cc}^- 
 \Tr\left[\overline H_a(v)H_b(v)\,\Aslash_{ba}\right]\cr
   &+{g \kappa_5 \over \Lambda_\chi} \CM_{dc}^+  
\Tr\left[\overline H_a(v)H_a(v)\,\Aslash_{cd}\gamma_5\right]
   +{g \kappa_6 \over \Lambda_\chi} \CM_{dc}^-  
     \Tr\left[\overline H_a(v)H_a(v)\,\Aslash_{cd}\right]\cr
 &+ {g \kappa_7 \over \Lambda_\chi}\CM^+_{cc} \Tr\left[\overline
H_a(v)\im v\cdot D_{ba} H_b(v)\right] + {g \kappa_8 \over
\Lambda_\chi}\CM^+_{ac} \Tr\left[\overline H_c(v)\im v\cdot D_{ba}
H_b(v)\right]\cr & + ... \cr}} where $\CM^{\pm} = \frac12(\xi m_q \xi
\pm \xi^\dagger m_q \xi^\dagger)$, and the ellipses denote operators
suppressed by additional powers of $M$, $m_s$, or derivatives.  The
\ord\ operators are exactly analogous to these, but with spin
preserving or spin violating Dirac structures (two Dirac structures
for each $\kappa$ term). Because chiral loops add a factor of $m_K^2$,
terms with additional pions are not relevent to the processes and
order we are working, so for our purposes, we may take $\xi \to 1$,
$\CM^+ \to m_q$, and $\CM^- \to 0$.  Of the $\kappa's$, only
$\kappa_1$ and $\kappa_5$ cannot then be absorbed by parameter and
field redefinitions; neither enters into the calulation of decay
constants. 
Note that the products $\kappa_i m_q$, rather than $\kappa_i$ and
$m_q$ separately, enter the Lagrangian, so there are no ambiguities
due to the definition of $m_s$.
The $\lambda$ terms may be accounted for by shifting
$v\cdot k \to v\cdot k - \delta$ in strangeness carrying heavy meson
propagators, with $\delta = M_{D_s} - M_D = M_{B_s} - M_B +
\O(\frac{\lbar^2}M) $ being the SU(3) mass splitting.

Omitting operators which are irrelevent to the calulations at hand
allows us to write the simplified Lagrangian
\eqn\lagrangem{\eqalign{
\CL_{M+m}&= -\Tr\left[\overline H_a(v)\im
v\cdot D_{ba} H_b(v)\right] + {\tilde g_{{\ol H}H} }\,\Tr\left[
\overline H_a(v)H_b(v)\,\Aslash_{ba}\gamma_5\right]\,\cr
 &+ {g \kappa_1 \over \Lambda_\chi} \CM_{ca}^+
  \Tr\left[\overline H_a(v)H_b(v)\,\Aslash_{bc}\gamma_5\right]
+{g \kappa_5 \over \Lambda_\chi} \CM_{dc}^+
\Tr\left[\overline H_a(v)H_a(v)\,\Aslash_{cd}\gamma_5\right]
       \cr }}
where
$$\gt = \cases{ \gt_{B^*} = g + \frac1M (g_1 + g_2) &for $B^* B^*$
coupling,\cr \gt_B = g + \frac1M (g_1 - g_2) &for $B^* B$
coupling,\cr}$$

\nref\chow{C-K Chow and M. Wise, \physrev{D48}{5202}{1993}\semi
L. Randall and M. Wise, \pl{B303}{135}{1993}}

The two new flavor violating terms appearing at this order correspond 
to spin symmetric and spin dependent axial coupling
renormalizations. As such, we expect them to affect heavy meson
interactions generically at \ord. For example, they will enter into
heavy quark and chiral symmetry violating corrections to semileptonic
$\ol B \to D$  and $\ol B_s \to D_s$ form factors, as well as to
hyperfine mass splittings. Previous work has neglected the effects of
these couplings\refs{\jenkins,\chow}.

The chiral representation of the left handed current $\ol\q_{aL}\Gamma\b$
proceeds similarly. The SU(3) preserving current is
    \eqn\currentm{\eqalign{
   J_{a(M)}^\lambda &=  \frac{i\alpha}2 (1 + \frac{\ro}M  )
          \Tr[\Gamma H_b(v)\xi^\dagger_{ba}] +
          \frac{i\alpha}2 \frac{\rt}{M}
         \Tr[\gamma^\mu \Gamma \gamma_\mu H_b(v)\xi^\dagger_{ba}] \cr
  &+ \frac{\omega_1}M \Tr[\Gamma H_b(v) \gamma_\mu D^\mu \xi^\dagger_{ba}]\cr
  &+ \frac{\omega_2}M \Tr[\gamma_\alpha \Gamma \gamma^\alpha H_b(v)
             \gamma_\mu D^\mu \xi^\dagger_{ba}] \cr
  &+ \frac{\omega_3}M \Tr[\Gamma v\cdot D_{cb} H_c(v) \xi^\dagger_{ba}]\cr
  &+ \frac{\omega_4}M \Tr[\gamma_\alpha \Gamma \gamma^\alpha
            v\cdot D_{cb} H_c(v) \xi^\dagger_{ba}]\cr
  &+ \frac{\omega_5}M \Tr[\Gamma H_b(v)  v\cdot D_{bc} \xi^\dagger_{ca}]\cr
  &+ \frac{\omega_6}M \Tr[\gamma_\alpha \Gamma \gamma^\alpha
            H_b(v) v\cdot D_{bc} \xi^\dagger_{ca}]\cr
  &+ \frac{i\alpha}{4 M} \Tr\left[ [\Gamma,\gamma_\mu]
            \im D^\mu_{cb} H_c(v)\xi^\dagger_{ba} \right] }}
where $\Gamma = \gamma^\lambda (1 - \gamma_5)$. Coefficients with even
subscripts multiply heavy quark spin violating operators.
The operators $\Tr[\Gamma \gamma_\mu H_b(v) D^\mu \xi^\dagger_{ba}]$
and $\Tr[\gamma_\alpha \Gamma \gamma^\alpha \gamma_\mu H_b(v)
D^\mu \xi^\dagger_{ba}]$ can be absorbed into $\omega_1$ and  $\omega_2$.
Terms proportional to the light quark mass matrix have been omitted because
their contributions are suppressed by ${m_K^2 \over \Lambda_\chi^2} $
compared to those in Eq.~\currentm. Potential terms involving the
axial current, such as $\Tr[\gamma_\alpha \Gamma \gamma^\alpha
              H_c(v) v\cdot A_{bc} \xi^\dagger_{ca}]$, can be
rewritten in terms of the listed operators by use of the identities\
$ \im D^\mu \xi^\dagger = - A^\mu \xi^\dagger$,
$ \im D^\mu \xi =  A^\mu \xi$.
\OMIT{identity
$ \im D^\mu \xi^\dagger = - A^\mu \xi^\dagger $.  More
generally, effects of operators such as $\Tr[\Gamma
\gamma_\mu H_c(v) (\xi^\dagger D^\mu \xi)_{cb} \xi^\dagger_{ba}$,
which are nonlinear in $\xi$ fields, can be absorbed by the
operators in Eq.~\currentm, as long as we are working to linear order
in the pion fields.}

The last term in Eq.~\currentm\ is given by VRI.  At tree level, its sole
effect is to turn parameters such as $v$ into physical quantities like
$\ol v = \frac{p_B}{M_B}$. For example, because of this VRI completion term,
the axial vector part of the current contributes
$\frac{\im \alpha}2 \Tr[\gamma^\lambda \gamma_5 \tilde H] =\im \alpha
(v + \frac{k}M)^\lambda =  \im \alpha \ol v^\lambda$,
while the vector part contributes $\frac{\im \alpha}2
\Tr[\gamma^\lambda \tilde H] = \im \alpha (\epsilon_v^\lambda
- v^\lambda \frac{k \cdot \epsilon_v}M )
= \im \alpha \epsilon_{\ol v}^\lambda$, where $\epsilon_v$
is the polarization
vector for the effective field $H_v$. The last relation follows from
transforming $\epsilon_v$ to $\epsilon_{\ol v}$ via a Lortentz
boost\vri.

We are now in a position to compute corrections to meson decays
at $\CO(\frac1M, m_s^0)$. At this order, the rate for $D^* \to D \pi$ is governed by
$g + \frac1{M_D} (g_1 - g_2)$ instead of $g$.
This is the quantity which is extracted from either the
$D^{*+}$ width\ref\accmor{The ACCMOR
Collaboration (S.~Barlag \etal), \pl{278}{1992}{480}} or the
$D^* \to D \gamma$ branching ratio\ref\radiative{J. Amundson, \etal,
\pl{296}{1992}{415}\semi
H-Y Cheng et al, \physrev{D47}{1993}{1030}\semi
P. Cho and H. Georgi, \pl{B296}{1992}{408}; erratum ibid {\bf
B300}(1993)410\semi 
 S. Stone, HEPSY-1-92, {\it Heavy Flavors},
A. Buras and H. Lindner, eds., World Scientific, Singapore (1992). }.
This leads to $$.1 < g^2 + \frac2{M_D}(g g_1 -g g_2) < .5.$$

The decay constants defined by
\eqn\matchconds{\eqalign{
    \langle 0\,|\,\ol\q_a\gamma^\mu\gamma^5\b\,|\,\ol B_a(p)\rangle
    &=\im f_{B_a}p^\mu,\cr
    \langle 0\,|\,\ol\q_a\gamma^\mu\b\,|\,\ol{B_a^*}(p,\epsilon)\rangle
    &=\im f_{B_a^*}\epsilon^\mu \cr }}
are altered only by current corrections
at this order:
\eqn\fd{
       \sqrt{M_D}f_{D} = \alpha [ 1 + \frac{\ro + 2 \rt}{ M_D}] }

\eqn\fdstar{
       \frac1{\sqrt{M_D}}f_{D^*} = \alpha [ 1 + \frac{\ro - 2\rt}{ M_D}] }

In principle, knowledge of $f_B$,$f_{B^*}$,$f_D$,$f_{D^*}$, and
$\Gamma(D^{*+} \to D \pi)$ would give $\alpha, \ro, \rt$ and
${\tilde g_B}$.  Only four couplings are determined because
the decay constants must satisfy
$$ \frac{f_{D^*}}{M_D f_D} - 1 = \frac{M_B}{M_D}(\frac{f_{B^*}}{M_B
f_B} - 1).$$

\nref\latt{A. Kronfeld, FNAL-CONF-93/277-T (hep-ph/9310220).}
A lattice calculation of the psuedo-scalar decay
constants\refs{\bernard,\latt},  finds ${f_B \over f_D} = .9 $
to $5\%$, from which we estimate $\ro + 2 \rt \approx -1.4$~GeV.

By matching a \ord\ calculation of $f_B$ to
the same calculation\neubert\ in the heavy quark effective field
theory (HQEFT)\ref\hqeft{N.~Isgur and M.~B.~Wise,
\pl{232}{1989}{113}; \pl {237}{1990}{527}\semi
E.~Eichten and B.~Hill, \pl{234}{1990}{511}\semi H.~Georgi,
\pl{240}{1990}{447}\semi A.~F.~Falk, B.~Grinstein and M.~Luke,
\np{357}{1991}{185}}, the current corrections can be related to matrix
elements in the effective theory.  The two HQEFT matrix elements $G_1$
and $G_2$, respectively related to $<0| \ol q \Gamma h \quad \ol h D^2
h | B>$ and $<0| \ol q \Gamma h \quad \ol h \sigma_{\mu\nu} G^{\mu\nu}
h | B>$, where $h$ is the effective heavy quark field and $
G^{\mu\nu}$ is the gluon field strength tensor, have been estimated
using QCD sum rules\neubert.  Matching gives $\ro = G_1 + 2 G_2 -
\frac{\lbar}6$ and $\rt = 2 G_2 - \frac{\lbar}6 $, while sum rule
estimates give $G_1 \approx -2.3$~GeV, $G_2 \approx .05$~Gev, and $\lbar
\approx .5 $~GeV.  This indicates $\rt \ll \ro$.  However, the same
author has recently called into question the large value of
$G_1$\virial. Moreover, an independent sum rule calculation
gives\ref\pball{P. Ball, \np{B421}{1994}{593}} $\rho_1=2\rho_2=-0.6$~GeV.

\INSERTFIG
{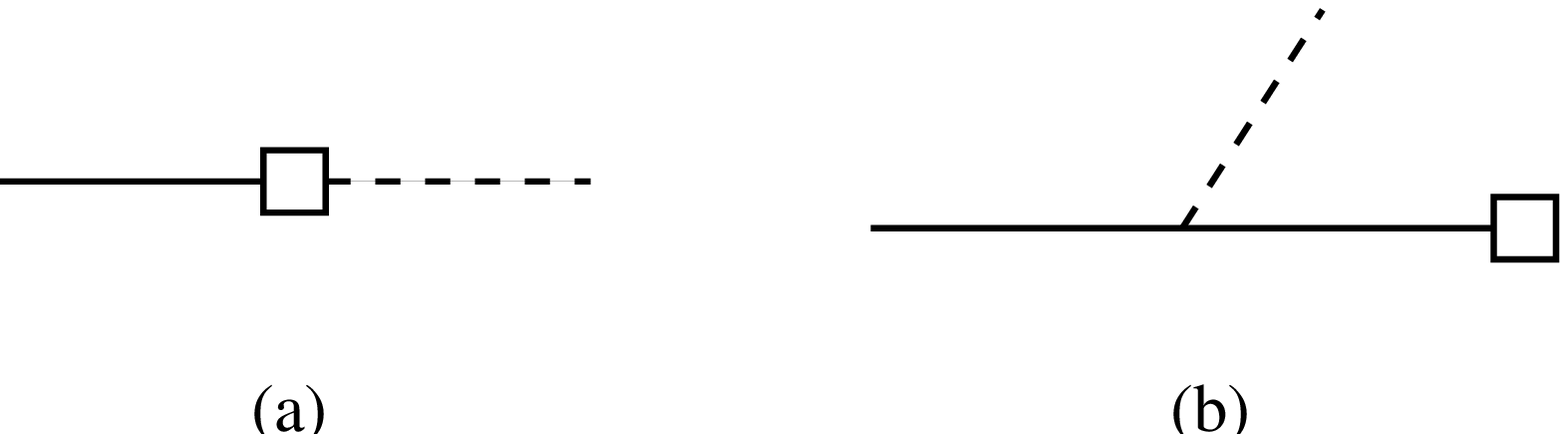}{\CAPTION{Figure 1.}{Tree level
diagrams for semileptonic $\ol B \to
\pi$. Solid lines represent heavy mesons, dashed lines
pseudogoldstone bosons. (a) The empty square indicates an
insertion of the \ordone\ heavy-light current. (b) The pole
amplitude with an insertion of the axial coupling followed
by annihilation via the current.
}}

We can also compute the \ord~, SU(3) symmetric corrections to 
semileptonic $\ol B \to K$.
%The SU(3) corrections to this process have been previously computed\btok.
%\ref\btok{A. Falk and B. Grinstein, SLAC-PUB-6237 and SSCL-Preprint-422
%(hep-ph/9306310).}.
The relevant matrix elements are
\eqn\formfactors{\eqalign{
    \langle \ol K(p_K)\,|\,\ol\s\gamma^\mu\b\,|\,\ol B(p_B)\rangle
    &=f_+\,(p_B+p_K)^\mu + f_-\,(p_B-p_K)^\mu\,,\cr
    \langle \ol K(p_K)\,|\,\ol\s\sigma^{\mu\nu}\b\,|\,\ol B(p_B)\rangle
    &= 2 \im h\,\[ p_K^\mu p_B^\nu - p_K^\nu p_B^\mu \]\,.\cr}}
Only the form factors $h(p_K\cdot p_B)$ and $f_+(p_K\cdot p_B)$
enter into the differential partial decay rate.

The operators which match onto the heavy-light currents above
are determined by equation \currentm\ to be
\eqn\matching{\eqalign{
    \O^\mu &={\im\over4}\alpha \,\{
       (1 + \frac{\ro - 2 \rt}{M_B})\Tr\left[\gamma^\mu
    H_b(v)(\xi^\dagger+\xi)_{ba}\right]\cr
    &+ (1 + \frac{\ro + 2 \rt}{M_B})\Tr\left[\gamma_5\gamma^\mu
    H_b(v)(\xi^\dagger-\xi)_{ba}\right]\}\cr
    \O^{\mu\nu} &={\im\over4}\alpha (1 + \frac{\ro}{M_B})\,
    \{\Tr\left[\sigma^{\mu\nu}H_b(v)(\xi^\dagger+\xi)_{ba}\right]\cr
    &+\Tr\left[\gamma_5\sigma^{\mu\nu}H_b(v)(\xi^\dagger-\xi)_{ba}\right]
    \} \,.\cr}}

The derivative suppressed terms in the current have been ignored.  They
give no contribution, at tree level, to $f_B$, while  for $\ol B \to \pi
l \nu$, they are down by~${k_\pi \over \Lambda_\chi}$.

The form factors implied by Fig.~(1) and Eq.~\matching\ are
\eqn\fplus{\eqalign{
           f_+ &=-{\alpha\over 2 \sqrt{M_B} f_\pi}
           \left[{M_B - v\cdot k \over v\cdot k- \Delta}
          (1 + \frac{\ro - 2 \rt}{M_B})\gt_B \right.\cr
       &\qquad\qquad\quad \left.+ (1 + \frac{\ro + 2 \rt}{M_B})\right]\cr
        &=-{1\over 2 f_\pi} \left[{\gt_Bf_{B^*}\over v\cdot k- \Delta}
        -{\gt_Bf_{B^*}\over M_B}+f_B\right]
 }}
and
\eqn\hformfactor{\eqalign{
            h &=-{\alpha \over 2 \sqrt{M_B} f_\pi}
           \left[{1 \over v\cdot k-\Delta}
          (1 + \frac{\ro}{M_B})\gt_B \right] \cr
           &=-{1\over 2f_\pi M_B}\left[{\gt_Bf_{B^*}\over v\cdot k- \Delta}
                (1+{2\rho_2\over M_B})\right],
}}
where $k$ is the pion momentum. The analogous formula for $D$ decays
apply with the substitution $M_B,f_B \to M_D,f_D$.  When expressed in
terms of physical quantities the expression for the pole part of
$f_+^{(D)}$ is not modified at \ord\ref\burgneub{G.~Burdman \etal,
Phys. Rev. D49 (1994) 2331.},
\eqn\fplussoft{ f_+^{(D)} = -{\gt_D f_{D^*} \over 2
f_\pi} {1 \over v\cdot k - \Delta }~,}
\nref\iw{N. Isgur and M. Wise, \physrev{D42}{1990}{2388}.}
but the relation between  $f_+^{(D)}$ and $h$ is\iw,
\eqn\fpluvshnew{
 f_+^{(D)} = M_D h^{(D)} (1 -\frac{2\rt}{M_D})~.
}
To the extent that $\rt$ is small, we expect relations between, say,
 $\ol B \to \ol K \mu^+ \mu^-$ and $\ol B \to \pi \mu \nu$ form factors to
be dominated by SU(3) rather than heavy quark corrections. In
principle, measurements of $f_+^{(B)}$, $f_+^{(D)}$ and either
$h^{(B)}$ or $h^{(D)}$, coupled with a precise measurement of $D^* \to
D \pi$, would determine all six unknown constants $\alpha$, $\ro$,
$\rt$, $g$, $g_1$ and $g_2$. 

Now consider the SU(3) breaking current, linear in $m_q$:
\eqn\currents{\eqalign{
   J_{a(m)}^\lambda &=
   \frac{i\alpha}2 (\frac{\eta_0}{\Lambda_\chi} + \frac{\eta_1}M)
    \Tr[\Gamma H_c(v) ]\CM^+_{cb} \xi^\dagger_{ba}
   +\frac{i\alpha}2 \frac{\eta_2}M
    \Tr[\gamma^\mu \Gamma \gamma_\mu H_c(v) ] \CM^+_{cb}\xi^\dagger_{ba} \cr
 &+ \frac{i\alpha}2 (\frac{\eta_3}{\Lambda_\chi} + \frac{\eta_4}M)
     \Tr[\Gamma H_b(v) \xi^\dagger_{ba} ] \CM^+_{cc}
  + \frac{i\alpha}2 \frac{\eta_5}M
\Tr[\gamma^\mu \Gamma \gamma_\mu H_b(v) \xi^\dagger_{ba} ]
  \CM^+_{cc}\cr
  &+ \frac{i\alpha}2 (\frac{\eta_6}{\Lambda_\chi} + \frac{\eta_7}M)
     \Tr[\Gamma H_c(v) \gamma_5] \CM^-_{cb} \xi^\dagger_{ba} 
  + \frac{i\alpha}2 \frac{\eta_8}M
\Tr[\gamma^\mu \Gamma \gamma_\mu H_c(v)  \gamma_5]
  \CM^-_{cb} \xi^\dagger_{ba} \cr
   &+ \frac{i\alpha}2 (\frac{\eta_9}{\Lambda_\chi} + \frac{\eta_{10}}M)
     \Tr[\Gamma H_b(v) \xi^\dagger_{ba} \gamma_5] \CM^-_{cc}
  + \frac{i\alpha}2 \frac{\eta_{11}}M
\Tr[\gamma^\mu \Gamma \gamma_\mu H_b(v) \xi^\dagger_{ba} \gamma_5]
  \CM^-_{cc}\cr }}
For $\xi \to 1$, the last two lines vanish, while the second line,
which is $SU(3)$ symmetric, can be absorbed by redefinitions of
$\alpha, \ro$, and $\rt$. Of the three remaining parameters
relevent at $\CO(\frac1M, m_s)$,  only $\eta_0$ contributes
to the pole part of $D \to K l \nu$ form factors,
so we can learn about it by looking at SU(3) violations 
among the form factors $f_+$ in the heavy quark limit.

The SU(3) loop corrections to these processes have been previously computed\btok,
so we will present only the counterterm contributions. In the Lagrangian
\lagrangem, $\kappa_1$ enters into decays involving both kaons and etas, while
$\kappa_5$ enters exclusively into decays involving etas. Since the 
SU(3) properties of the $\eta_0$ and $\kappa_1$ terms differ, $\eta_0$
can be extracted, in principle, from the following two form factors:
\eqn\btokaon{
f_+^{D \to K l \overline \nu} =-{g \alpha \sqrt{M_B} \over 2 f_\pi v\cdot k}
        [1 +  {m_s \over \Lambda_\chi}(\eta_0 + \kappa_1)],}
and
\eqn\btokaons{
f_+^{D_s \to K l \overline \nu} =-{g \alpha \sqrt{M_B} \over 2 f_\pi v\cdot k}
        [1 +  {m_s \over \Lambda_\chi}\kappa_1].}
These relations, augmented by the known loop corrections, relate $\eta_0$
to decay constant corrections.

\smallskip
III. LOOP CORRECTIONS
\smallskip

The leading SU(3) and heavy quark symmetry violating contributions to
$f_B$ come from one loop diagrams involving both virtual kaons and
\orlm\ heavy quark violating contact terms (with coefficients
$\ro,\rt,g_1,g_2$), and arise from
nonanalytic dependence on the strange quark mass $m_s$.  Since this
nonanalytic dependence is only generated by chiral loops, we can
compute chiral and heavy quark symmetry violation, at leading order,
in terms of the six parameters $\alpha$, $\ro$, $\rt$, $g$, $g_1$ and
$g_2$. The formally sub-leading counterterms $\eta_0,\eta_1$, and $\eta_2$ are
added to absorb the scheme and subtraction scale dependence of the loop
calculation, and to make the expansion more reliable numerically.

The nonanalytic $\mu$ dependent parts of
several integrals which arise during the loop computation have
been compiled using dimensional regularization\btok. Here we 
retain the complete expression, including a pole contained 
in $\overline \Delta = \frac2\epsilon -\gamma + \ln(4\pi) +1$.
We will use
\eqn\ioneitwo{\eqalign{
    &\im\pintegral {1\over p^2-m^2}={1\over16\pi^2}I_1(m)\,,\cr
    &\im\pintegral {1\over (p^2-m^2)(p\cdot v-\Delta)}=
    {1\over16\pi^2}{1\over\Delta}I_2(m,\Delta)\,,\cr}}
and
\eqn\jmunu{\eqalign{
    J^{\mu\nu}(m,\Delta)&=\im\pintegral{p^\mu p^\nu\over(p^2-m^2)
    (p\cdot v-\Delta)}\cr
    &={1\over16\pi^2}\Delta\left[J_1(m,\Delta)g^{\mu\nu}
    +J_2(m,\Delta)v^\mu v^\nu\right]\,,\cr}}
where
\eqn\idefs{\eqalign{
    I_1(m)&=m^2\ln(m^2/\mu^2) - m^2 \dbar \,,\cr
    I_2(m,\Delta)&=-2\Delta^2\ln(m^2/\mu^2)-4\Delta^2F(m/\Delta)
          +2 \Delta^2 + 2 \dbar \Delta^2\,.\cr}}
\eqn\fdef{F(x)=\left\{\eqalign{
    &\sqrt{1-x^2}\;\tanh^{-1}\sqrt{1-x^2}\,,\cr
    &-\sqrt{x^2-1}\;\tan^{-1}\sqrt{x^2-1}\,,\cr}
    \qquad\eqalign{&|x|\le1\cr &|x|\ge 1\cr}\right.}
\eqn\joneandtwo{\eqalign{
    J_1(m,\Delta)&=(-m^2+\frac23\Delta^2)\ln(m^2/\mu^2)+\frac43
    (\Delta^2-m^2)F(m/\Delta)\cr
     &- \frac23 \Delta^2 (1 + \dbar) +\frac13 m^2 (2 +3 \dbar)\,,\cr
    J_2(m,\Delta)&=(2m^2-\frac83\Delta^2)\ln(m^2/\mu^2)-\frac43
    (4\Delta^2-m^2)F(m/\Delta)\cr
       &+ \frac83 \Delta^2 (1 + \dbar) -\frac23 m^2 (1+3\dbar)\,.\cr}}
We choose  $\overline \Delta =0$ for our calculations. The analytic terms from
these integrals, which were dropped from the SU(3) loop corrections in 
reference \btok, are particularly easily recovered with this scheme.

The demonstration that higher derivative operators give no \ord~contribution
requires only the nonanalytic terms of several integrals which are easily 
derived from the above integrals. The needed integrals are
%\eqn\kmunu{\eqalign{
%    K^{\mu\nu}(m,\Delta)&=\im\pintegral{p^\mu p^\nu\over
%    (p^2-m^2)(p\cdot v-\Delta)^2}\cr
%     &= ...  .\cr}}

\eqn\lmunulambda{\eqalign{
  L^{\mu\nu\lambda}(m,\Delta)
                 &=\im\pintegral{p^\mu p^\nu p^\lambda \over(p^2-m^2)
    (p\cdot v-\Delta)}\cr
    &= \frac13 g^{(\mu\nu}v^{\lambda)}\[ L_1 - L_2 \]
  + v^\mu v^\nu v^\lambda \[ 2 L_2 - L_1 \], \cr }}
where
\eqn\lonedef{
\eqalign{
 L_1 &=\im\pintegral{p^2 (v \cdot p) \over(p^2-m^2)(p\cdot v-\Delta)}\cr
   &= \frac{m^4}{16 \pi^2}\ln(m^2/\mu^2) - {2 \Delta^2 m^2
\over 16 \pi^2}\[\ln(m^2/\mu^2) + 2 F(m/\Delta)\], \cr }}
\eqn\ltwodef{\eqalign{
 L_2  &= \im\pintegral{(v \cdot p)^3 \over(p^2-m^2)(p\cdot v-\Delta)}\cr
   &= \frac14 \frac{m^4}{16 \pi^2}\ln(m^2/\mu^2)  +
     {\Delta^2 \over 16 \pi^2}\[(m^2 - 2 \Delta^2)
      \ln(m^2/\mu^2) - 4 \Delta^2 F(m/\Delta) \],\cr   }}
and
\eqn\ltwomunulambda{\eqalign{
  {\partial \over \partial \Delta} &L^{\mu\nu\lambda}(m,\Delta)
     =\im\pintegral{p^\mu p^\nu p^\lambda \over(p^2-m^2)
    (p\cdot v-\Delta)^2 }\cr
   &= g^{(\mu\nu}v^{\lambda)}{2 \Delta \over 16 \pi^2}
  \left[ (\frac43 \Delta^2 - m^2)\ln(\frac{m^2}{\mu^2})
    + \frac23 F(\frac m\Delta) {2 m^4 + 4 \Delta^4 -7 m^2 \Delta^2
    \over  \Delta^2 - m^2 } \right]\cr
  &+  v^\mu v^\nu v^\lambda {2 \Delta \over 16 \pi^2}
    \left[ (4 m^2 - 8 \Delta^2) \ln(\frac{m^2}{\mu^2})
     + 2 F(\frac{m}{\Delta}){8 m^2 \Delta^2 - m^4 - 8 \Delta^4 \over
\Delta^2 - m^2} \right], \cr }}
as well as
\eqn\mmunualpbet{\eqalign{
    M^{\mu\nu\alpha\beta}(m,\Delta)&=\im\pintegral{p^\mu p^\nu p^\alpha
       p^\beta \over(p^2-m^2) (p\cdot v-\Delta)^2}\cr
 &= \frac1{30} g^{(\mu\nu}g^{\alpha\beta)}\[M_1(m,\Delta)
-2 M_2(m,\Delta) + M_3(m,\Delta)\] \cr
       & +\frac1{15} g^{(\mu\nu}v^\alpha v^{\beta)}
         \[-M_1(m,\Delta) + 7 M_2(m,\Delta) -6 M_3(m,\Delta)\]  \cr
        &  + \frac15 v^\mu v^\nu v^\alpha v^\beta
           \[M_1(m,\Delta) - 12 M_2(m,\Delta) + 16 M_3(m,\Delta)\] , \cr }}
where
\eqn\mone{\eqalign{
    M_1(m,\Delta)& = \im\pintegral{p^4 \over(p^2-m^2) (p\cdot v-\Delta)^2}\cr
                 &= {-m^4 \over 16 \pi^2 }\[ 2 \ln(\frac{m^2}{\mu^2})
         + {4 \Delta^2 \over \Delta^2 - m^2}F(\frac{m}{\Delta}) \]\cr}}

\eqn\mtwo{\eqalign{
    M_2(m,\Delta)& = \im\pintegral{p^2 (v\cdot p)^2 \over(p^2-m^2)
       (p\cdot v-\Delta)^2}\cr
        &= {1 \over 16 \pi^2 }(m^4 - 6 m^2 \Delta^2)\ln(\frac{m^2}{\mu^2})
        + {4 \Delta^2 m^2 \over 16 \pi^2} F(\frac{m}{\Delta})
         {3 \Delta^2 - 2 m^2 \over m^2 - \Delta^2}, \cr }}
\eqn\mthree{\eqalign{
    M_3(m,\Delta)&=
       \im\pintegral{(v\cdot p)^4 \over(p^2-m^2) (p\cdot v-\Delta)^2}\cr
                 &= {1 \over 16 \pi^2 }(\frac14 m^4 + 3 m^2 \Delta^2
     - 10 \Delta^4)\ln(\frac{m^2}{\mu^2}) + {4 \Delta^4 \over 16 \pi^2}
     { 4 m^2 - 5 \Delta^2 \over \Delta^2 - m^2} F(\frac{m}{\Delta}) \cr }}
and ( ) means symmetrization, {\it e.g.},
\eqn\symet{\eqalign{
 g^{(\mu\nu}v^\alpha v^{\beta)} &= g^{\mu\nu}v^\alpha v^\beta +
       g^{\mu\alpha}v^\nu v^\beta + g^{\mu\beta} v^\nu v^\alpha \cr
           & +g^{\nu\alpha}v^\mu v^\beta
         + g^{\nu\beta} v^\mu v^\alpha + g^{\alpha\beta} v^\mu v^\nu
          \cr }}
The entire nonanalytic parts of these integrals
have been retained, even though we will only use the leading
terms. The
analytic parts, including the divergent $\epsilon$-pole, have been
discarded. It is  worth pointing out that there are no singularities
at $m^2 = \Delta^2$ in the $M_i$: the apparent pole is cancelled by a
zero in the function~$F$.
\OMIT{
%\eqn\mmunu{\eqalign{
%   M^{\mu\nu}(m,\Delta)&=\im\pintegral{p^2 p^\mup^\nu\over(p^2-m^2)
%    (p\cdot v-\Delta)^2}\cr
%    &= g^{\mu\nu}\left[ {m^2(3 \Delta^2 - m^2) \over 24 \pi^2}
%       - {4 m^4\over 3}\right]\ln(m^2/\mu^2) \cr
%    &+ v^\mu v^\nu \left[{m^2 ( m^2 - 12 \Delta^2) \over 24 \pi^2}
%     + {16m^4 \over 3}\right] \ln(m^2/\mu^2).\cr}}
}

\INSERTFIG
{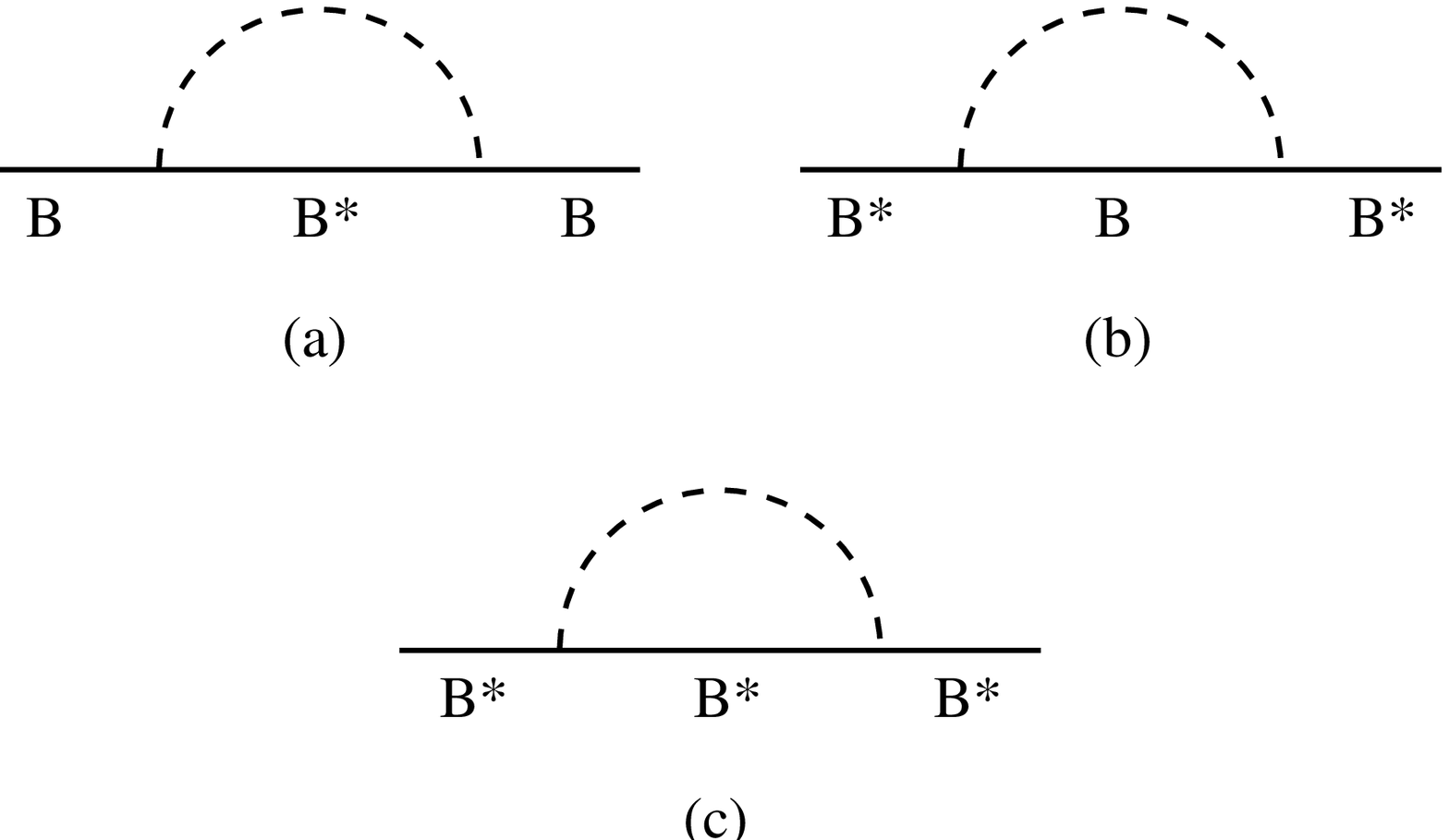}
\CAPTION{Figure 2.}{Wavefunction renormalization diagrams for
(a) pseudoscalar heavy mesons, and (b,c) vector mesons.
}

The diagrams of Fig.~(2) contribute to wavefunction renormalization.
The graph of Fig.~(2a), after summing over intermediate states and ignoring
the pion mass, is
\eqn\borz{
     { 6 \im \tilde g_B^2 \over 16 \pi^2 f^2}
    \[ J(m_K, \dfour + \delta - v \cdot k ) + \frac16 J(m_\eta,\dfour
    - \vk )\] }
for the $B^0$, and
\eqn\bsrz{
     {6 \im \tilde g_B^2 \over 16 \pi^2 f^2}
 \[ 2 J(m_K, \dfour - \vk) + \frac23 J(m_\eta,\dfour + \delta
     - \vk)\] }
for the $B_s$.

For the vector mesons, there are two graphs, Figs.~(2b) and~(2c).
Together, they give (we drop the $v^\mu v^\nu$ terms)
\eqn\bostarrz{\eqalign{
     &{-2 \im \over 16 \pi^2 f^2}
    \big[2 \tilde g_{B^*}^2 J(m_K, \dfour +\delta -\vk) + \frac13
     \tilde g_{B^*}^2 J(m_\eta,\dfour -\vk)
          \cr
    &+
    \tilde g_B^2 J(m_K, \dmthree +\delta -\vk) + \frac16 \tilde g_B^2
    J(m_\eta,\dmthree -\vk)\big] g^{\mu\nu}
            \cr } }

for the ${B^0}^*$, and
\eqn\bsstarrz{\eqalign{
     &{-2 \im \over 16 \pi^2 f^2}
    \big[ 4 \tilde g_{B^*}^2 J(m_K, \dfour - \vk) + \frac43 \tilde
     g_{B^*}^2 J(m_\eta,\dfour +\delta -\vk) \cr
    &+
    2 \tilde g_B^2 J(m_K, \dmthree - \vk) + \frac23
    \tilde g_B^2 J(m_\eta,\dmthree +\delta -\vk)\big] g^{\mu\nu}\cr }  }
for the $B_s^*$, where $J(m,x) = x J_1(m,x)$.

Differentiating the above self energies with respect to
$2 \vk$ and evaluating on-shell gives the wavefunction
renormalizations $Z$. Expanding the couplings $\tilde g$,
applying the Gell-Mann-Okubo formula $m^2_\eta = \frac43 m_K^2$, 
and noting $\Delta \sim \frac1M$, yields
\eqn\zbo{\eqalign{
       Z_{B^0} = &1
      - \frac{11}3 \Lz g^2 + \frac38 \Lfp  \cr
         &\quad- \frac{22}{3} \LMz
        (g_1 g - g_2g ) }}

\eqn\zbs{\eqalign{
       Z_{B_s} = &1
      - \frac{26}3 \Lzst g^2  + \frac34 \Lfm \cr
         &\quad- \frac{52 }{3} \LMzst
        (g_1g - g_2g ) }}

\eqn\zbostar{\eqalign{
       Z_{{B^0}^*} = &1
      - \frac{11}3 \Lz g^2 + \frac18 \Lfm \cr
         &\quad- \frac{22}{9} \LMz
        (3 g_1g + g_2g ) }}

\eqn\zbsstar{\eqalign{
       Z_{B_s^*} =&1
      - \frac{26}3 \Lzst g^2 +\frac14 \Lfp \cr
         &\quad- \frac{52 }{9} \LMzst
        (3 g_1g + g_2g ) }}
The chiral log terms are the leading SU(3) corrections because they
are logarithmically enhanced in the chiral $m_K \to 0$ limit. Whether
$F({m_K / \Delta})$ is similarly enhanced depends on how we take the
combined heavy quark and chiral limit -- the behavior of the ratio
${m_K / \Delta}$ affects the answer we get, including the coefficient
of the chiral log. A consistent but non-unique scheme is to take $m_K
\to 0$ holding ${m_K^2 / \Lambda_\chi \Delta}$ fixed.  Rather than
choosing a particular scheme, we retain nonanalytic terms such as
$F({m / (\Delta + \delta)})$. Note that while
$F(m/x)\to -\frac\pi2 m/x\to\infty$ as $x\to0$, the function involved
is always of the form $x^2 F(m/x)$, which vanishes in the limit.
Therefore, we do not drop terms of order $\Delta^2$ or $\delta^2$ when
they appear in the combination $(\Delta\pm\delta)^2$ as a factor
multiplying $ F({m / (\Delta\pm\delta)})$, even if they are formally
of order $1/M^2$ and $m_K^4$, respectively. 

\INSERTFIG
{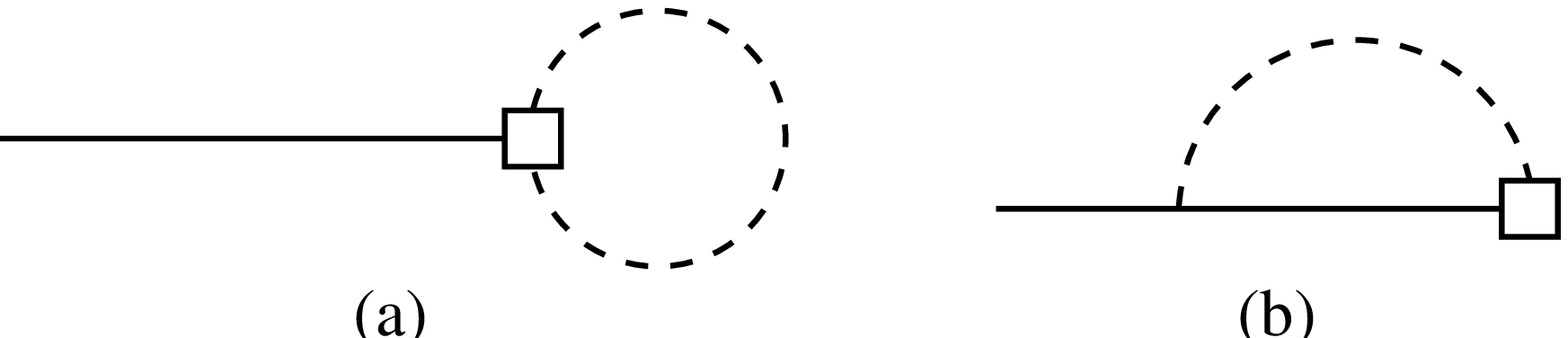}
\CAPTION{Figure 3.}{Diagrams contributing to decay constants by
modifying the current vertex.
}

Vertex corrections arise from the diagrams in Fig.~(3).
The graph of Fig.~(3b) vanishes at \orlm. Fig.~(3a) gives,
\eqn\vertexbo{
  {\im \alpha v^\lambda \over 32 \pi^2 f^2}
   (1 + \frac{\ro + 2 \rt}{M})\[ I_1(m_K^2) + \frac16 I_1(m_\eta^2)\] }
for the $B^0$ meson,
\eqn\vertexbs{
  {\im \alpha v^\lambda \over 32 \pi^2 f^2}
   (1 + \frac{\ro + 2 \rt}{M})\[ 2 I_1(m_K^2) + \frac23 I_1(m_\eta^2)\] }
for the $B_s$,
\eqn\vertexbostar{
  {- \im \alpha \epsilon^{\lambda} \over 32 \pi^2 f^2}
   (1 + \frac{\ro - 2 \rt}{M})\[ I_1(m_K^2) + \frac16 I_1(m_\eta^2)\] }
for the ${B^0}^*$ meson, and
\eqn\vertexbostar{
  {- \im \alpha \epsilon^{\lambda} \over 32 \pi^2 f^2}
   (1 + \frac{\ro -2 \rt}{M})\[ 2 I_1(m_K^2) + \frac23 I_1(m_\eta^2)\] }
for the $B_s^*$.

\INSERTFIG
{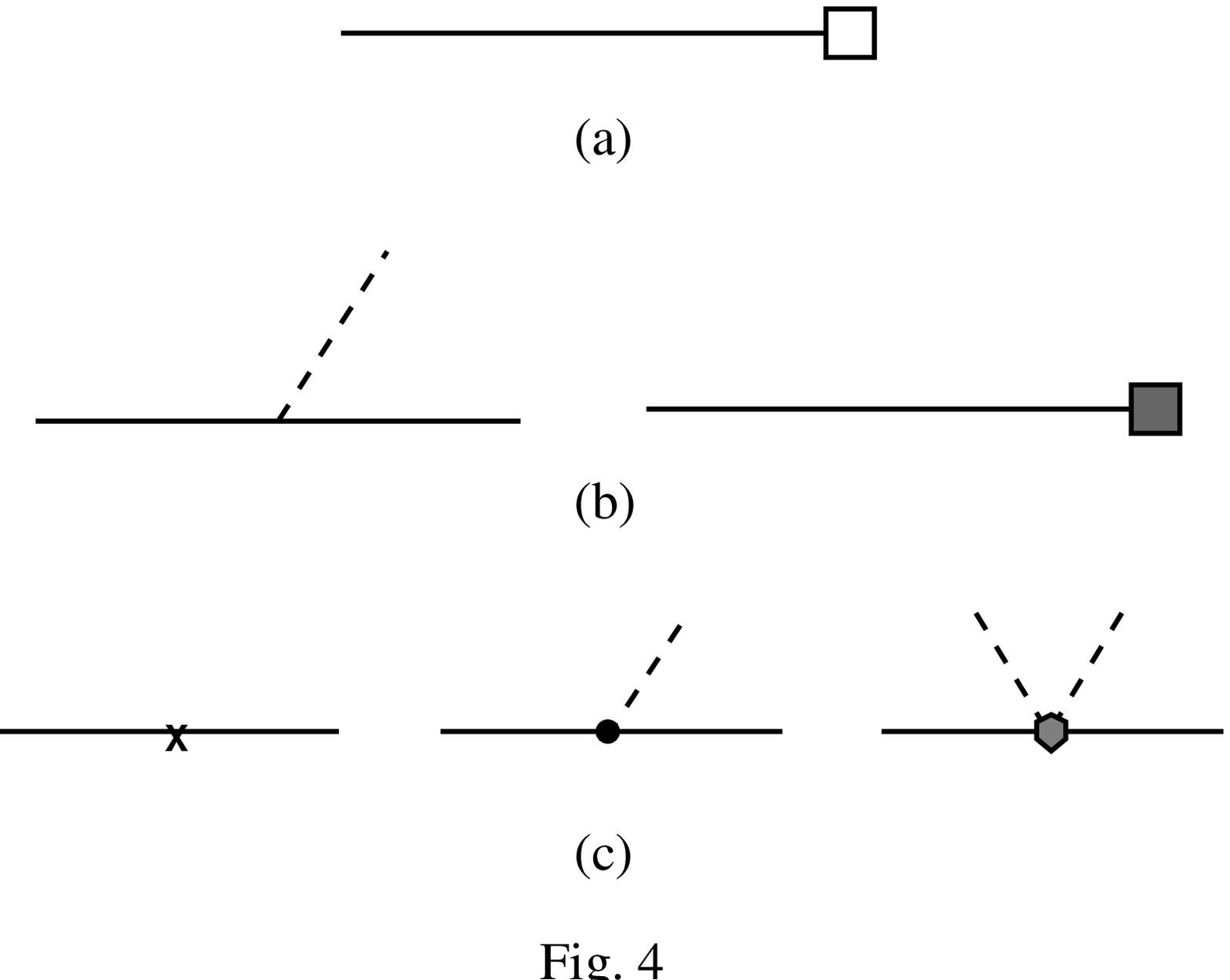}
\CAPTION{Figure 4.}{Vertices relevant to one loop diagrams by
which higher derivative operators contribute to decay
constants. (a) The \ordone\ current vertex is independent of
loop or external momenta $\sim 1$. (b) Both the \ordone\
axial coupling and the derivative suppressed
\ord\ current correction (shaded square) are linear in momentum,
$\sim p^\mu$. (c) Derivative operators in the Lagrangian can
modify the two-point function (cross), the axial coupling
(solid dot), or the vector coupling (shaded dot) at \ord.  
All are quadratic in momentum, $\sim p^\mu p^\nu$.  }

At $\O(\frac{k}M)$, Fig.~(3b) no longer vanishes because terms in
equations \origlagranem\ and \currentm\ with extra derivatives
contribute. To justify our claim that such terms are suppressed by
${\Delta \over M} \sim \frac1{M^2}$ rather than ${m_K \over M}$, we
explicitly compute the contribution of the ${\alpha \over M}$ VRI
completion term of the current (the last term in Eq.~\currentm) to the
graph in Fig.~(3b). For an incoming $B^0$ turning into a $B_s^* K^0$
loop, the graph gives \eqn\vriloop{
    {3 \im \alpha g \Delta v^\lambda \over M_B 64 \pi^4 f_K^2}
             J_1(m_K,\dfour)}
which is suppressed by $\frac1{M^2}$. The other higher derivative
operators are similarly suppressed.

\INSERTFIG
{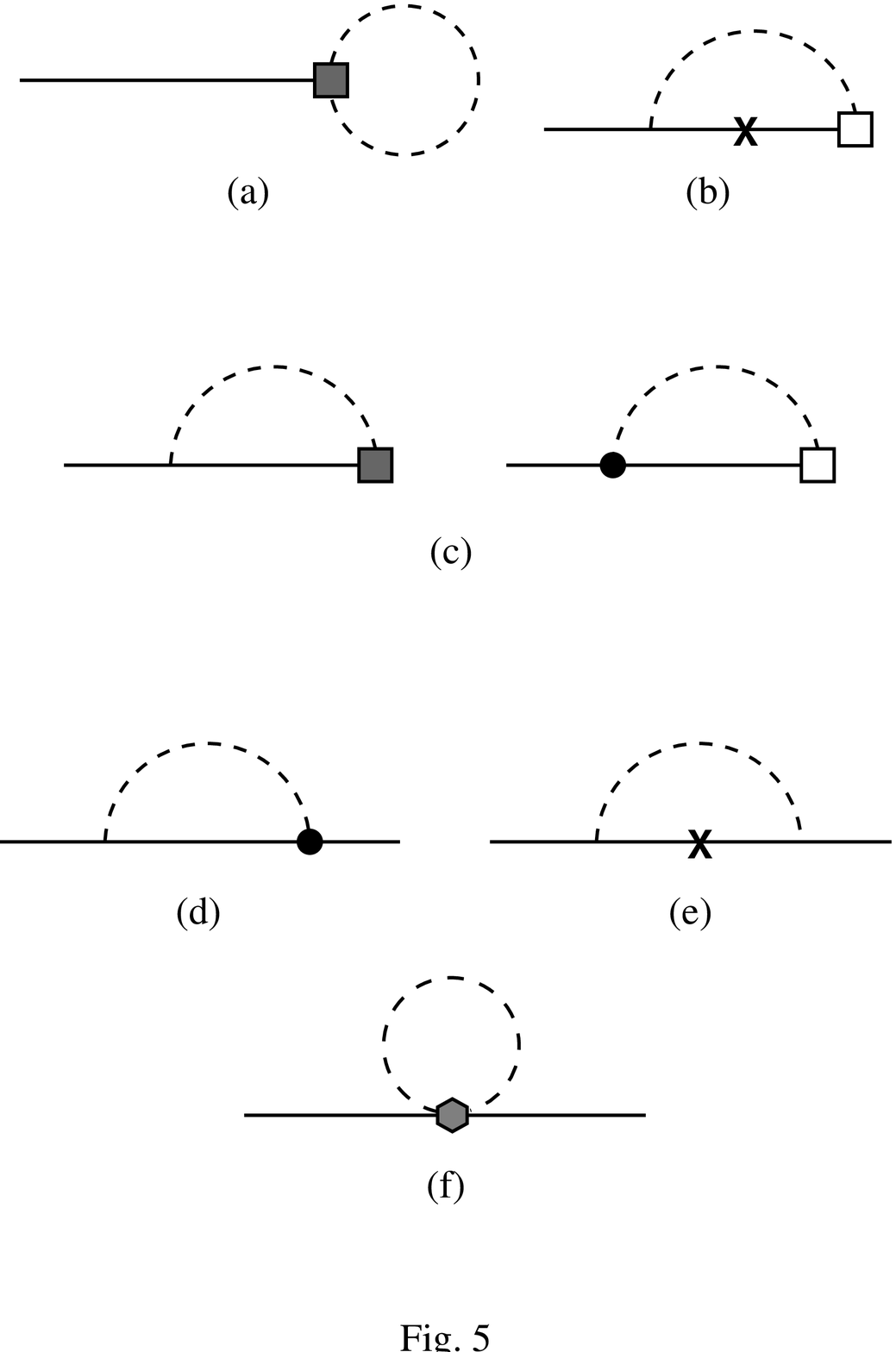}
\CAPTION{Figure 5.}{Graphs through which higher derivative
operators in the current or Lagrangian can contribute to
decay constants. Diagrams (a), (b) and~(c) represent corrections
to the current vertex. The diagrams of (d), (e) and (f) are
wavefunction renormalizations.
}

We can outline why this is the case by classifying the graphs involving
higher derivative operators according to their associated integrals. The
relevant vertices
are grouped in Fig.~(4).  The leading, $\O(1)$, part of
the current shown in Fig.~(4a) has no powers of momentum in its vertex,
independent of whether the incoming meson is spin zero or one. The 
$\O(M^0)$  axial coupling and the \ord\ current vertex in Fig.~(4b) are linear
in momentum, while the higher derivative Lagrangian insertion and
pion couplings in fig. (4c) are quadratic in momentum. Since we are free to
take the residual momentum of the incoming, on-shell, meson to be zero,
only loop momentum will appear in the graphs of Fig.~(5) (except for
one power of residual momentum needed for the wavefunction renormalization
graphs of Figs.~(5d),~(5e) and~(5f)).

The vertex graph in Fig.~(5a) involves the integral
\eqn\fivea{
   \pintegral {p^\mu \over p^2-m^2} }
which trivially vanishes. Vertex corrections involving derivative
suppressed Lagrangian insertions, as in Fig.~(5b), are proportional to
$\frac1M{\partial \over \partial
\Delta}L^{\mu\nu\lambda}(m,\Delta)$.  Since we treat $F(m/\Delta)$ as
$\CO(1)$, one can see from Eq.~\ltwomunulambda\ that these corrections
are \orms.  Both vertex loops of Fig.~(5c) involve $\frac1M
J^{\mu\nu}(m,\Delta)$, which is also subleading.

For wavefunction renormalization, we consider the diagram of Fig.~(5d),
containing
an axial coupling correction,
which involves $ \frac1M L^{\mu\nu\lambda}(m,\Delta + v\cdot k)$, where
$k$ is the heavy meson's residual momentum.  From Eq.~\lmunulambda, we see
that the term proportional to $v\cdot k$ contains an additional power
of $\Delta$, so it is down by \orms.  The loop graph in Fig.~(5e)
contains a factor of $\frac1M M^{\mu\nu\alpha\beta}$.  Again, the terms
with at least one factor of $v\cdot k$ automatically come with at least
one factor
of $\Delta$. For either graph, replacing one of the loop momenta
in the numerator with the residual momentum fails to alter the $\frac1M$
suppression.  The last graph, Fig.~(5f), can contribute to
wavefunction renormalizaion at \ord\ only if the four-point
coupling is linear in the external momentum, in which
case it is also linear in the pion momentum, causing
the loop integral to vanish.
Thus, none of the higher derivative operators contribute
to the leading heavy quark and chiral symmetry violating terms.

The final results for the decay constants, valid to
$\O({\lbar \over M},{m_K \over M})$, are found by combining
the wavefunction, vertex, and counterterm corrections:
\eqn\fbo{\eqalign{
       \sqrt{M_{B^0}}&f_{B^0} = \alpha \biggl[ 1 + \frac{\ro + 2 \rt}{M}
      - \frac{11}6 \L (g^2 + \frac13)  \biggr.\cr
      &- \frac13 (\frac{11}3 + \ln \frac43)\mg
    + \frac3{16} \Lfp \cr
         &- \frac{11}{6} \LM
        ( \ro (g^2 + \frac13) +2 \rt (g^2 + \frac13)
        +2 g g_1 - 2 g g_2 )  \cr
    &\biggl. -\frac13 (\frac{11}3 + \ln \frac43) \mM 
    ( (\ro +2 \rt) g^2 +2 g g_1 - 2 g g_2)\biggr]\cr
                  }}

\eqn\fbs{\eqalign{
     \sqrt{M_{B_s}}&f_{B_s} = \alpha \biggl[ 1 + \frac{\ro + 2 \rt}{M}
      - \frac{13}3 \L (g^2 + \frac13)  \biggr.\cr
          &-\frac43 (\frac{13}6 + \ln \frac43) \mg
        + \frac38 \Lfm \cr
         &- \frac{13}{3} \LM
        ( \ro (g^2 +\frac13) +2 \rt (g^2 +\frac13)
        + 2 g g_1 - 2 g g_2 ) \cr
     &-\frac43 (\frac{13}6 + \ln \frac43)
        \mM ( (\ro +2 \rt) g^2 +2 g g_1 - 2 g g_2)\cr
   & \biggl.+ m_s ({\eta_1 + 2 \eta_2 \over M} 
	+ {\eta_0 \over \Lambda_\chi})\biggr]
    \cr }}

\eqn\fbostar{\eqalign{
      \frac1{\sqrt{M_{{B^0}^*}}}&f_{{B^0}^*}
           = \alpha \biggl[ 1 + \frac{\ro -2  \rt}{M}
      - \frac{11}6 \L (g^2 + \frac13)\biggr.\cr
     &- \frac13 (\frac{11}3 + \ln \frac43) \mg
    + \frac1{16} \Lfm  \cr
            & - \frac{11}{6} \LM
        ( \ro (g^2 +\frac13) - 2 \rt (g^2 +\frac13)
        + 2 g g_1 + \frac23  g g_2 ) \cr
    &\biggl.-\frac13 (\frac{11}3 + \ln \frac43) \mM 
     ( (\ro -2 \rt) g^2 +2 g g_1 + \frac23 g g_2)\biggr]\cr
          }}

\eqn\fbsstar{\eqalign{
      \frac1{\sqrt{M_{B_s^*}}}&f_{B_s^*} = 
	\alpha \biggl[ 1 + \frac{\ro -2  \rt}{M}
      - \frac{13}3 \L (g^2 + \frac13)\biggr.\cr
       &-\frac43 (\frac{13}6 + \ln \frac43) \mg
      + \frac18 \Lfp  \cr
            & - \frac{13}{3} \LM
        ( \ro (g^2 + \frac13) -2 \rt (g^2 + \frac13)
        + 2 g g_1 + \frac23  g g_2 ) \cr
       &-\frac43 (\frac{13}6 + \ln \frac43) \mM 
   ( (\ro -2 \rt) g^2 +2 g g_1 + \frac23 g g_2)\cr
    &\biggl. + m_s ({\eta_1 - 2 \eta_2 \over M} 
               + {\eta_0 \over \Lambda_\chi})\biggr]\cr
   }}
Equations \fbo\ to \fbsstar\ constitute the main results of this paper,
and agree with Ref.~\fchiralcalc\ in the infinite mass limit.

The physical decay constants and masses are independent of the choice
of renormalization point $\mu$. On the right hand side of Eqs.~\fbo\
to~\fbsstar\ the explicit dependence on the renormalization point
$\mu$ is cancelled by implicit dependence in the parameters $\alpha$,
$g_i$, $\rho_i$ and $\eta_i$. To see this one must recall that we have
absorbed counterterms with $\mu$ and analytic $m_K^2$ dependence into
these parameters.

It is worth pointing out that if we ignore analytic and
${\cal O}({1\over M^2}, {\delta^2/M})$ terms, the decay constants can be
written in the simpler form 
\eqn\fbosimple{\eqalign{
   \sqrt{M_{B^0}}f_{B^0} = &\alpha ( 1 + \frac{\ro + 2 \rt}{M})
    \Bigg\{1  - \frac{11}6 \L (\hat g^2 + \frac13)  \cr
      & + \frac3{16} \Lfph\Bigg\} ,\cr}}
and analogous formulas for the other decay constants.
Except for the last term, this form is simply 
the SU(3) correction to the infinite mass limit,
but with $\alpha ( 1 + \frac{\ro + 2 \rt}{M})$ and $\hat g$
replacing $\alpha$ and $g$. For $f_B$, $\hat g$ is simply
$\gt_B$, but for $f_{B^*}$, we need $\hat g = g + \frac{g_1
+ \frac13 g_2}{M} \ne \gt_{B^*} $. 
The constants $g_1$ and~$g_2$ do not simply enter through the
Lagrangian parameters $g_B$ and $g_{B^*}$; their contributions to
$f_{B^*}$ must be computed from the diagrams.
 
 From these equations, it follows that the ratio of decay constants $R_1$ is
\eqn\ratiovalue{\eqalign{
       R_1 =& 1
  - {9 g^2 \delta (\ln\frac{m_K^2}{\mu^2}-\frac13) \over 8 \pi^2 f_K^2}
                 (\Delta^{(B)} - \Delta^{(D)}) \cr
       &+ {3 g^2 \over 8 \pi^2 f_K^2 }
         \left[  -(\Ddp{B})^2  \fbplus
       + 2 (\Ddm{B})^2 \fbminus \right.\cr
        &\qquad\qquad\qquad +\left. (\Ddp{D})^2  \fcplus
        - 2 (\Ddm{D})^2  \fcminus \right]\cr
          &- 5\Lzrone (\frac1{M_B} - \frac1{M_D} )
             g (g_1 - g_2) 
      + m_s (\eta_1 +2 \eta_2) (\frac1{M_B} - \frac1{M_D} ) \cr } }
This quantity is relevant to the extraction from $B - \ol B$ mixing
of the Cabbibo-Kobayashi-Maskawa angle $V_{ts}$\ratio.

We can estimate the size of corrections to $R_1$.
Taking $\mu = 1$~GeV and inserting known masses and
constants gives
\eqn\rnum{
 R_1 -1=-0.11 g^2 -0.06\GeV^{-1} g (g_1 - g_2) -0.10 (\eta_1 +2 \eta_2)}
The first term contributes $-6\%$ to $R_1-1$ for $g^2= 0.5$.

Information about $g_1$ and $g_2$
may be gleaned by studying $\bar B_s \to D_s e \nu$ decays,
hyperfine mass splittings, or $\bar B \to \pi l \nu$. The presence
of $g_1,g_2$ in the Lagrangian means they will contribute
to any process involving $B$ or $D$ mesons which
recieve corrections due to pion loops. The counterterms
$\eta_1$ and $\eta_2$ enter at one loop into $\bar B \to \pi l \nu$
form factors.  It remains to be seen if predictive power persists 
in this system at $\CO(\frac1M, m_s)$. However, the presence of a large 
number of form factors (both the pole and constant parts) for the set of 
semileptonic decays $\{B,B_s,D,D_s\} \to \{K,\pi,\eta\}$ looks promising,
since many of the counterterms listed here may be eliminated by redefinitions 
of fields and coupling constants.

\bigskip
IV. CONCLUSIONS
\bigskip

We have derived the most general form of the heavy quark chiral Lagrangian
and the heavy to light current
to linear order in the $SU(3)$ vector and axial fields and
to \ord. Four new unknown parameters $\ro, \rt, g_1, g_2$ arise at this
order. Additional parameters, needed for terms suppressed by one
derivative,  have been presented, but the contribution of these
terms to the decay constants are found to be $\O(\frac1{M^2})$.
$\CO(m_s)$ counterterms in the Lagrangian and current have also been presented,
leading to three new parameters 
$\eta_0,\eta_1$, and $\eta_2$ which enter into decay constant relations.

The $\CO(\frac1M, m_s^0)$ heavy quark symmetry corrections to 
$\ol B \to \pi l \ol \nu$ and its flavor related decays
are computed in terms of the four unknown parameters 
$\ro, \rt, g_1, g_2$. Derivative
terms are suppressed by $k_\pi/\Lambda_\chi$.  SU(3) splittings
are examined at tree level, giving rise to relations which allow the extraction
of $\eta_0$ from $D \to K l \overline \nu$ and
$D_s \to K l \overline \nu$ form factors.

The $\CO(\frac1M, m_s^0)$ corrections to $f_B$ involve only
two of the parameters, which may to be related to HQEFT matrix
elements by comparison to the heavy quark effective theory.

Finally, the logarithmically enhanced contributions to heavy 
meson decay constants violating both chiral SU(3) and heavy quark 
symmetries can be
expressed in terms of two leading order constants $\alpha$ and $g$,
and four new parameters $\ro, \rt, g_1, g_2$.  Three unenhanced
counterterms $\eta_0,\eta_1$ and $\eta_2$ also enter.
The ratio $R_1 =
\frac{f_{B_s}}{f_B}/\frac{f_{D_s}}{f_D}$ is expressed in terms of
two unknown axial parameters and two counterterms.  

In principle, measurements of $f_+^{(B)}$, $f_+^{(D)}$ and either
$h^{(B)}$ or $h^{(D)}$, coupled with a precise measurement of $D^* \to
D \pi$, would determine six unknown constants $\alpha$, $\ro$,
$\rt$, $g$, $g_1$ and $g_2$.  Measurements of SU(3) violation
in $f_+$ or $h$ determine, in addition, $\eta_0$.
The observed decay constant $f_{D_s}$,
the potentially measurable $f_D$, $f_B$ and $f_{B_s}$, and the
inaccessible decay constants $f_{D^*}, f_{D_s^*}, f_{B^*}$ and
$f_{B_s^*}$ are then fixed in terms of two parameters
$\eta_1$ and $\eta_2$.  Predictions of inaccessible quantities
may be of use for lattice or hadron model computations. For example,
neither quenched lattice nor non-relativistic quark models include the
nonanalytic terms computed here, and should be augmented by our
results. 

An obvious extension of this work is the computation of 
$\CO(\frac1M,m_s)$
corrections to processes such as semileptonic $\ol B \to D$ and $\ol B
\to K K$, and $\ol B \to \pi$.  These processes depend on many of the same 
parameters presented in this work.
Combinations of such quantities sensitive only to simultaneous
violations of chiral and heavy quark symmetry may be 
amenable to fruitful examination.

\bigskip
\qquad  ACKNOWLEDGEMENTS
\bigskip

A treatment of $\frac1M$ corrections has been performed independently
by T. Kurimoto and N. Kitazawa in DPNU 93-37, OS-GE 38-93. We wish to
thank them for sharing their insights with us. G.~B. thanks Adam Falk,
Mike Luke, and Aneesh Manohar for illuminating discussions, and
acknowledges the support of the Department of Energy, under contract
DOE--FG03--90ER40546.  B.~G.~is supported in part by the Alfred P.~Sloan
Foundation and by the Department of Energy under contract
DE--FG05--92ER--40722.  This work was initiated at the Superconducting
Super Collider Laboratory, with support from the Department of Energy
under contract DE--AC35--89ER40486.

\listrefs
\vfil\eject
\INSERTCAP{\bigskip
\qquad \qquad  {\rm Figure Captions}:
\smallskip
Fig.\ 1.\quad  Tree level diagrams for semileptonic $\ol B \to
\pi$. Solid lines represent heavy mesons, dashed lines
pseudogoldstone bosons. (a) The empty square indicates an
insertion of the \ordone\ heavy-light current. (b) The pole
amplitude with an insertion of the axial coupling followed
by annihilation via the current.
\smallskip
Fig.\ 2.\quad Wavefunction renormalization diagrams for
(a) pseudoscalar heavy mesons, and (b,c) vector mesons.
\smallskip
Fig.\ 3.\quad Diagrams contributing to decay constants by
modifying the current vertex.
\smallskip
Fig.\ 4.\quad Vertices relevant to one loop diagrams by
which higher derivative operators contribute to decay
constants. (a) The \ordone\ current vertex is independent of
loop or external momenta $\sim 1$. (b) Both the \ordone\
axial coupling and the derivative suppressed
\ord\ current correction (shaded square) are linear in momentum,
$\sim p^\mu$. (c) Derivative operators in the Lagrangian can
modify the two-point function (cross), or the axial coupling
(solid dot) at \ord.  Both are quadratic in momentum, $\sim
p^\mu p^\nu$.
\smallskip
Fig.\ 5.\quad Graphs through which higher derivative
operators in the current or Lagrangian can contribute to
decay constants. Diagrams (a), (b) and~(c) represent corrections
to the current vertex. The diagrams of (d) and (e) are
wavefunction renormalizations.
\vfil\eject}
\bye